\documentclass[12pt]{article}
\usepackage{amsfonts}
\usepackage{latexsym,amsmath,amssymb}
\usepackage{graphics,epstopdf}
\usepackage[pdftex]{graphicx}
\numberwithin{equation}{section}

\newtheorem{thm}{Theorem}[section]

\numberwithin{equation}{section}

\begin{document}

\bigskip

\bigskip

\begin{center}
{\Large \textbf{
B$\acute{e}$zier curves and surfaces based on  modified Bernstein polynomials
 }}

\bigskip

\textbf{Khalid Khan,}$^{1)}$ \textbf{D. K. Lobiyal}$^{1)}$ and  \textbf{Adem kilicman}$^{2)}$

$^{1)}$School of Computer and System Sciences, SC \& SS, J.N.U., New Delhi-110067., India%
\\[0pt]khalidga1517@gmail.com; dklobiyal@gmail.com\\
$^{2)}$Department of Mathematics,
Faculty of Science, University Putra
Malaysia, Malaysia \\[0pt]akilicman@putra.upm.edu.my \\[0pt]

\bigskip

\bigskip

\textbf{Abstract}
\end{center}

\parindent=8mm {\footnotesize {In this paper, we use the blending functions of Bernstein polynomials with shifted knots for construction of B$\acute{e}$zier curves and surfaces. We study  the nature of degree elevation and degree reduction for B$\acute{e}$zier Bernstein functions with shifted knots for $t \in \big[\frac{\alpha}{n+\beta} , \frac{n+\alpha}{n+\beta}\big]$. Parametric curves are represented using these modified Bernstein basis and the concept of total positivity is applied to investigate the shape properties of the curve. We get B$\acute{e}$zier curve defined on $[0,1]$  when we set the parameter $\alpha,\beta$ to the value $0.$
 We also present a de Casteljau algorithm to compute Bernstein B$\acute{e}$zier curves and surfaces with shifted knots. The new curves have some properties similar to B$\acute{e}$zier curves. Furthermore, some fundamental properties for Bernstein B$\acute{e}$zier curves and surfaces are discussed.

\bigskip


{\footnotesize \emph{MSC: primary 65D17; secondary 41A10, 41A25, 41A36.}: \newline

\bigskip

\section{Introduction}

\parindent=8mm It was  S.N. Bernstein \cite{brn} in 1912, who first introduced his famous operators $%
B_{n}: $ $C[0,1]\rightarrow C[0,1]$ defined for any $n\in \mathbb{N}$
and for any function $f\in C[0,1]$  where $ C[0,1]$  denote the set of all continuous
functions on $[0,1]$ which is equipped with sup-norm $\Vert .\Vert _{C[0,1]}$%
\begin{equation}\label{e1.1}
B_{n}(f;x)=\sum\limits_{k=0}^{n}\left(
\begin{array}{c}
n \\
k%
\end{array}%
\right) x^{k}(1-x)^{n-k}f\biggl{(}\frac{k}{n}\biggl{)},~~x\in \lbrack 0,1].
\end{equation}
 and named it Bernstein polynomials to prove the Weierstrass
theorem \cite{pp}.

Bernstein showed that if $f\in C[0,1]$, then $B_{n}(f;x)\rightrightarrows f(x)$
where $"\rightrightarrows "$ represents the uniform convergence. One can
find a detailed monograph about the Bernstein polynomials in \cite{lor}.

Later it was found that Bernstein polynomials possess many remarkable properties and has various applications in areas such as approximation theory \cite{pp}, numerical analysis,
computer-aided geometric design, and solutions of differential equations due to its fine properties of approximation \cite{hp}.

In  computer aided geometric design (CAGD), Bernstein polynomials and its variants are used in order to preserve the shape of the curves or surfaces. One of the most important curve in CAGD \cite{thomas} is the classical B$\acute{e}$zier curve \cite{Bezier} constructed with the help of Bernstein basis functions. Other works related to different generalization of Bernstein polynomials and bezier curves and surfaces can be found in \cite{cetin1,cetin2,farouki,khalid, khalid1,lp,mka1,mur8,ma1,pl1,pl,hp,mahmudov1,sofia}

\parindent=8mm In 1968 Stancu \cite{sta} showed that the polynomials
\begin{equation}\label{e1.2}
\bigl{(}P^{(\alpha,\beta)}_nf\bigl{)}(x)=\sum\limits_{k=0}^{n}{n\choose k}x^k(1-x)^{n-k}f\biggl{(}\frac{k+\alpha}{n+\beta}\biggl{)}
\end{equation}
converge to continuous function $f(x)$ uniformly in [0,1] for each real $%
\alpha ,\beta $ such that $0\leq \alpha \leq \beta $. The polynomials (\ref{e1.2})
are called as a Bernstein-Stancu polynomials.\newline

\parindent=8mm In 2010, Gadjiev and Gorhanalizadeh \cite{gad} introduced the
following construction of Bernstein-Stancu type polynomials with shifted
knots:

\begin{equation}\label{e1.3}
S_{n,\alpha,\beta}(f;x)=\biggl{(}\frac{n+\beta_2}{n}\biggl{)}^n\sum\limits_{k=0}^{n}{n\choose k}\biggl{(}x-\frac{\alpha_2}{n+\beta_2}\biggl{)}^k\biggl{(}\frac{n+\alpha_2}{n+\beta_2}-x\biggl{)}^{n-k}f\biggl{(}\frac{k+\alpha_1}{n+\beta_1}\biggl{)}
\end{equation}
where $\frac{\alpha_2}{n+\beta_2}\leq x\leq\frac{n+\alpha_2}{n+\beta_2}$ and
$\alpha_k,\beta_k~(k=1,2)$ are positive real numbers provided $%
0\leq\alpha_1\leq\alpha_2\leq\beta_1\leq\beta_2$. It is clear that for $%
\alpha_2=\beta_2=0$, then polynomials (\ref{e1.3}) turn into the Bernstein-Stancu
polynomials (\ref{e1.2}) and if $\alpha_1=\alpha_2=\beta_1=\beta_2=0$ then these
polynomials turn into the classical Bernstein polynomials.\\

 In recent years, generalization of the B$\acute{e}$zier curve with shape parameters has received continuous attention.
 Several authors were concerned with the problem of changing the shape of curves and surfaces, while keeping the
control polygon unchanged and thus they generalized the B$\acute{e}$zier curves in \cite{khalid, khalid1,wcq,hp}. \\

%

The outline of this paper is as follow: Section $2$ introduces a modified Bernstein functions with shifted knots  $G_{n,\alpha,\beta}^k$ and their Properties. Section $3$ introduces degree elevation and degree reduction properties for these modified Bernstein functions. Section $3.2$ introduces a de Casteljau algorithm  for $G_{n,\alpha,\beta}^k$. In Section $4$ we define a tensor
product patch based on algorithm $1$ and its geometric properties as well as a degree
elevation technique are investigated. Furthermore tensor product of B$\acute{e}$zier surfaces on $\big[\frac{\alpha}{n+\beta} , \frac{n+\alpha}{n+\beta}\big] \times \big[\frac{\alpha}{n+\beta} , \frac{n+\alpha}{n+\beta}\big]$  for Bernstein polynomials with shifted knots are introduced and its properties that is inherited from the univariate case are being discussed.\\

In next section, we construct basis functions with shifted knots with the help of (\ref{e1.3}).

\section{Bernstein functions with shifted knots}

The Bernstein functions with shifted knots is defined
as follows

\begin{equation}\label{e2.1}
G_{n,\alpha,\beta}^k(t)={n\choose k}~~\biggl{(}\frac{n+\beta}{n}\biggl{)}^n~~\biggl{(}t-\frac{\alpha}{n+\beta}\biggl{)}^k\biggl{(}\frac{n+\alpha}{n+\beta}-t\biggl{)}^{n-k}
\end{equation}
where $\frac{\alpha}{n+\beta}\leq t\leq\frac{n+\alpha}{n+\beta}$ and
$\alpha,\beta~$ are positive real numbers provided $%
0\leq\alpha\leq\beta$.

\subsection{Properties of the Bernstein functions with shifted knots}

\begin{thm}
   The Bernstein functions with shifted knots  possess the following properties:\\

(1.) Non-negativity: $G_{n,\alpha,\beta}^k~(t)\geq 0~~~~$
 $k = 0, 1, . . . , n,~~~ t \in \big[\frac{\alpha}{n+\beta}, \frac{n+\alpha}{n+\beta}\big].$\\

(2.)Partition of unity:

\begin{equation*}
\sum\limits_{k=0}^{n} G_{n,\alpha,\beta}^k~(t)= 1, ~~~~\text{for every}~~ t \in \big[\frac{\alpha}{n+\beta}, \frac{n+\alpha}{n+\beta}\big]..
\end{equation*}\\

(3.)  End-point interpolation property holds:
\begin{equation*}
G_{n,\alpha,\beta}^k~\big(\frac{\alpha}{n+\beta}\big)=\left\{
\begin{array}{ll}
1,~~~~~\mbox{if $k=0$ } &  \\
&  \\
0,~~~~~~~~~~\mbox{$ k \neq 0$} &
\end{array}%
\right.
\end{equation*}

\begin{equation*}
G_{n,\alpha,\beta}^k~\big(\frac{n+\alpha}{n+\beta}\big)=\left\{
\begin{array}{ll}
1,~~~~~~~~~\mbox{if $k=n$ } &  \\
&  \\
0,~~~~~~~~~~\mbox{$ k \neq n $} &
\end{array}%
\right.
\end{equation*}\\
 clearly both sided end point interpolation property holds.\\

(4.) Reducibility: when $\alpha=\beta=0$ formula $(\ref{e2.1})$ reduces to the classical Bernstein bases on $[0,1]$.\\
\end{thm}

\textbf{Proof:} All these property can be deduced easily from  equation (\ref{e2.1}).\\

\begin{figure*}[htb!]
\begin{center}
\includegraphics[height=6cm, width=8cm]{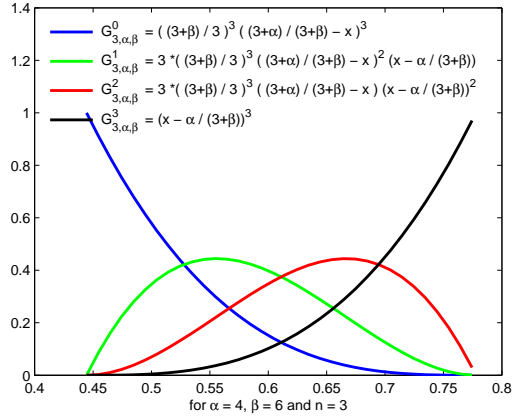}
\end{center}
\caption{`Cubic Bezier blending functions with shifted knots'}\label{f1}
\end{figure*}

\begin{figure*}[htb!]
\begin{center}
\includegraphics[height=6cm, width=8cm]{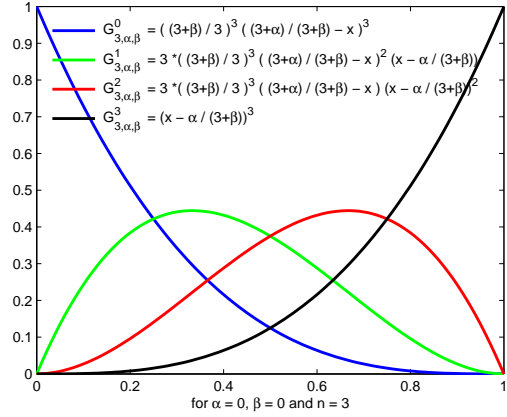}
\end{center}
\caption{`Cubic Bezier blending functions'}\label{f2}
\end{figure*}

Fig. $\ref{f1}$ shows the modified Bernstein basis functions of degree $3$ with shifted knots for $\alpha=4,\beta=6.$  Here we can observe that sum of blending  fuctions is always unity and also satisfies end point interpolation property. In case $\alpha=\beta=0,$  it turns out to be classical Bernstein basis on $[0,1]$ which is shown in Fig. $\ref{f2}.$\\

Apart from the basic properties above, the Bernstein functions with shifted knots also satisfy some
recurrence relations, as for the classical Bernstein basis.\\

\section {Degree elevation and reduction for Bernstein functions with shifted knots}
Technique of degree elevation has been used to increase the flexibility of a given curve.
A degree elevation algorithm calculates a new set of control points by choosing a convex combination
of the old set of control points which retains the old end points. For this purpose, the identities (\ref{e3.1}),(\ref{e3.2}) and Theorem (\ref{t3.1}) are useful.\\

\textbf{Degree elevation}

\begin{equation}\label{e3.1}
\bigg(\frac{n+\alpha}{n+\beta}-t\bigg)G_{n,\alpha,\beta}^k~(t)= \bigg(\frac{n+1-k}{n+1}\bigg)\bigg(\frac{n}{n+\beta}\bigg)G_{n+1,\alpha,\beta}^k~(t)
\end{equation}

and
\begin{equation}\label{e3.2}
  \bigg(t-\frac{\alpha}{n+\beta}\bigg)G_{n,\alpha,\beta}^k=\bigg(\frac{n}{n+\beta}\bigg)\bigg(\frac{k+1}{n+1}\bigg)G_{n+1,\alpha,\beta}^{k+1}
\end{equation}

\textbf{Proof:}

Consider

 \begin{equation*}
\bigg(\frac{n+\alpha}{n+\beta}-t\bigg) G_{n,\alpha,\beta}^k~=\bigg(\frac{n+\alpha}{n+\beta}-t\bigg)\{ {n\choose k}~~\biggl{(}\frac{n+\beta}{n}\biggl{)}^n~~\biggl{(}t-\frac{\alpha}{n+\beta}\biggl{)}^k\biggl{(}\frac{n+\alpha}{n+\beta}-t\biggl{)}^{n-k}   \}
\end{equation*}

 \begin{equation*}
\bigg(\frac{n+\alpha}{n+\beta}-t\bigg) G_{n,\alpha,\beta}^k~=\{ \frac{{n\choose k}}{{{n+1}\choose k}}{{n+1}\choose k}~~\biggl{(}\frac{n+\beta}{n}\biggl{)}^n~~\biggl{(}t-\frac{\alpha}{n+\beta}\biggl{)}^k\biggl{(}\frac{n+\alpha}{n+\beta}-t\biggl{)}^{n-k+1}   \}
\end{equation*}

 \begin{equation*}
\bigg(\frac{n+\alpha}{n+\beta}-t\bigg) G_{n,\alpha,\beta}^k~=\{ \frac{{n\choose k}}{{{n+1}\choose k}}~~\biggl{(}\frac{n}{n+\beta}\biggl{)}   \}G_{n+1,\alpha,\beta}^{k}~(t)
\end{equation*}

 \begin{equation*}
\bigg(\frac{n+\alpha}{n+\beta}-t\bigg) G_{n,\alpha,\beta}^k~=\bigg(\frac{n+1-k}{n+1}\bigg)\biggl{(}\frac{n}{n+\beta}\biggl{)}   G_{n+1,\alpha,\beta}^{k}~(t).
\end{equation*}

Similarly for

\begin{equation*}
  \bigg(t-\frac{\alpha}{n+\beta}\bigg)G_{n,\alpha,\beta}^k=\bigg(\frac{n}{n+\beta}\bigg)\bigg(\frac{k+1}{n+1}\bigg)G_{n+1,\alpha,\beta}^{k+1}
\end{equation*}

\begin{equation*}
  \bigg(t-\frac{\alpha}{n+\beta}\bigg)G_{n,\alpha,\beta}^k=\bigg(x-\frac{\alpha}{n+\beta}\bigg)\{ {n\choose k}~~\biggl{(}\frac{n+\beta}{n}\biggl{)}^n~~\biggl{(}t-\frac{\alpha}{n+\beta}\biggl{)}^k\biggl{(}\frac{n+\alpha}{n+\beta}-t\biggl{)}^{n-k}   \}
\end{equation*}

\begin{align*}
 \bigg(t-\frac{\alpha}{n+\beta}\bigg)G_{n,\alpha,\beta}^k &= \bigg(x-\frac{\alpha}{n+\beta}\bigg)\{ {n\choose k}~~\biggl{(}\frac{n+\beta}{n}\biggl{)}^n~~\biggl{(}t-\frac{\alpha}{n+\beta}\biggl{)}^k\biggl{(}\frac{n+\alpha}{n+\beta}-t\biggl{)}^{n-k}   \} \\
 &= {n\choose k}~~\biggl{(}\frac{n+\beta}{n}\biggl{)}^n~~\biggl{(}t-\frac{\alpha}{n+\beta}\biggl{)}^{k+1}\biggl{(}\frac{n+\alpha}{n+\beta}-t\biggl{)}^{n-k}\\
&=\frac{{n\choose k}}{{n+1\choose k+1}}~~{n+1\choose k+1}\biggl{(}\frac{n+\beta}{n}\biggl{)}^n~~\biggl{(}t-\frac{\alpha}{n+\beta}\biggl{)}^{k+1}\biggl{(}\frac{n+\alpha}{n+\beta}-t\biggl{)}^{n-k}\\
&=\bigg(\frac{n}{n+\beta}\bigg)\bigg(\frac{k+1}{n+1}\bigg)G_{n+1,\alpha,\beta}^{k+1}
\end{align*}

 \begin{thm}\label{t3.1}

  Each Bernstein functions with shifted knots of degree n is a linear combination of two Bernstein functions with shifted knots of degree $n+1:$
\begin{equation}\label{e3.3}
G_{n,\alpha,\beta}^k~(t)= \bigg(\frac{n+1-k}{n+1}\bigg)G_{n+1,\alpha,\beta}^k~(t)+\bigg(\frac{k+1}{n+1}\bigg)G_{n+1,\alpha,\beta}^{k+1}~(t)
\end{equation}
where

 $\frac{\alpha}{n+\beta}\leq t \leq\frac{n+\alpha}{n+\beta}$ and
$\alpha,\beta~$ are positive real numbers satisfying $%
0\leq\alpha\leq\beta$.

\end{thm}

\textbf{Proof:}

 \begin{equation*}
\bigg(\frac{n}{n+\beta}\bigg)~~G_{n,\alpha,\beta}^k~(t)= ~~G_{n,\alpha,\beta}^k~ {\bigg( \frac{n+\alpha}{n+\beta}-t+\{t-\frac{\alpha}{n+\beta}\}\bigg) }
\end{equation*}

 \begin{equation*}
\bigg(\frac{n}{n+\beta}\bigg)G_{n,\alpha,\beta}^k~(t)=\bigg(\frac{n+\alpha}{n+\beta}-t\bigg) G_{n,\alpha,\beta}^k~+ \bigg(t-\frac{\alpha}{n+\beta}\bigg)G_{n,\alpha,\beta}^k
\end{equation*}

on using equation $(\ref{e3.1}),(\ref{e3.2}),$ we can easily get

\begin{equation*}
G_{n,\alpha,\beta}^k~(t)= \bigg(\frac{n+1-k}{n+1}\bigg)G_{n+1,\alpha,\beta}^k~(t)+\bigg(\frac{k+1}{n+1}\bigg)G_{n+1,\alpha,\beta}^{k+1}~(t)
\end{equation*}

 \begin{thm}\label{t3.2}

  Each Bernstein functions with shifted knots of degree n is a linear combination of two Bernstein functions with shifted knots of degree $n-1:$
\begin{equation}\label{e3.4}
G_{n,\alpha,\beta}^k~(t)= \frac{n+\beta}{n} \bigg(t-\frac{\alpha}{n+\beta}\bigg) G_{n-1,\alpha,\beta}^{k-1}~(t)+\frac{n+\beta}{n} \bigg( \frac{n+\alpha}{n+\beta}-t \bigg) G_{n-1,\alpha,\beta}^{k}~(t)
\end{equation}
where

 $\frac{\alpha}{n+\beta}\leq t \leq\frac{n+\alpha}{n+\beta}$ and
$\alpha,\beta~$ are positive real numbers satisfying $%
0\leq\alpha\leq\beta$.

\end{thm}

\textbf{Proof} On using  Pascal's type relation i.e  , we get

\begin{align*}
  G_{n,\alpha,\beta}^k~(t)& ={n\choose k}~~\biggl{(}\frac{n+\beta}{n}\biggl{)}^n~~\biggl{(}t-\frac{\alpha}{n+\beta}\biggl{)}^k\biggl{(}\frac{n+\alpha}{n+\beta}-t\biggl{)}^{n-k}\\
  &=\{{n-1\choose k-1}+{n-1\choose k}\}~~\biggl{(}\frac{n+\beta}{n}\biggl{)}^n~~\biggl{(}t-\frac{\alpha}{n+\beta}\biggl{)}^k\biggl{(}\frac{n+\alpha}{n+\beta}-t\biggl{)}^{n-k}\\
  &= {n-1\choose k-1} \biggl{(}\frac{n+\beta}{n}\biggl{)}^n~~\biggl{(}t-\frac{\alpha}{n+\beta}\biggl{)}^k\biggl{(}\frac{n+\alpha}{n+\beta}-t\biggl{)}^{n-k}\\
  ~~&+{n-1\choose k} \biggl{(}\frac{n+\beta}{n}\biggl{)}^n~~\biggl{(}t-\frac{\alpha}{n+\beta}\biggl{)}^k\biggl{(}\frac{n+\alpha}{n+\beta}-x\biggl{)}^{n-k}\\
  &=\frac{n+\beta}{n} \bigg(t-\frac{\alpha}{n+\beta}\bigg) G_{n-1,\alpha,\beta}^{k-1}~(t)+\frac{n+\beta}{n} \bigg( \frac{n+\alpha}{n+\beta}-t \bigg) G_{n-1,\alpha,\beta}^{k}~(t)
\end{align*}

\begin{thm} The end-point property of derivative:

\begin{equation}\label{e3.5}
{\bf{P^{\prime}}}(\frac{\alpha}{n+\beta})= (n+\beta)({\bf{P_1}}-{\bf{P_0}}){\bigg(\frac{n-1+\beta}{n-1}\bigg)}^{n-1}{\bigg( \frac{n-1+\alpha}{n-1+\beta}-\frac{\alpha}{n+\beta}\bigg)}^{n-1-k}
\end{equation}

  \begin{equation}\label{e3.6}
  {\bf{P^{\prime}}}(\frac{n+\alpha}{n+\beta})= (n+\beta)({\bf{P_n}}-{\bf{P_{n-1}}}){\bigg(\frac{n-1+\beta}{n-1}\bigg)}^{n-1}{\bigg( \frac{n+\alpha}{n+\beta}-\frac{\alpha}{n-1+\beta}\bigg)}^{n-1}
\end{equation}

i.e. Bernstein-B$\acute{e}$zier curves with shifted knots are tangent to fore-and-aft edges of its control polygon at end points.
\end{thm}
\textbf{Proof:}
Let

\begin{align*}
  {\bf{ P}}(t)&= \sum\limits_{k=0}^{n} {\bf{P_k}}~ G_{n,\alpha,\beta}^k(t)\\
  &=\sum\limits_{k=0}^{n} {\bf{P_k}}{n\choose k}~~\biggl{(}\frac{n+\beta}{n}\biggl{)}^n~~\biggl{(}t-\frac{\alpha}{n+\beta}\biggl{)}^k\biggl{(}\frac{n+\alpha}{n+\beta}-t\biggl{)}^{n-k} \\
&={\bf{V}}(t)
\end{align*}
 or
 $${\bf{P}}(t)~ = {\bf{V}}(t)$$\\
 then on differentiating both hand side with respect to `t', we have\\
 $${\bf{ P^\prime}}(t)~ = {\bf{V^\prime}}(t).$$\\
 Let $$A_{k}^{n}(t)=~{n\choose k}~~\biggl{(}\frac{n+\beta}{n}\biggl{)}^n~~\biggl{(}t-\frac{\alpha}{n+\beta}\biggl{)}^k\biggl{(}\frac{n+\alpha}{n+\beta}-t\biggl{)}^{n-k}$$ then

$${\bf{V}}(t)=\sum\limits_{k=0}^{n}~{\bf{P_k}} A_{k}^{n}(t)$$\\

\begin{align*}
{(A_{k}^{n}(t)}^{\prime}&={n\choose k}~~\biggl{(}\frac{n+\beta}{n}\biggl{)}^n k~~\biggl{(}t-\frac{\alpha}{n+\beta}\biggl{)}^{k-1}\biggl{(}\frac{n+\alpha}{n+\beta}-t\biggl{)}^{n-k}\\
&-{n\choose k}~~\biggl{(}\frac{n+\beta}{n}\biggl{)}^n ~~\biggl{(}t-\frac{\alpha}{n+\beta}\biggl{)}^{k} (n-k)\biggl{(}\frac{n+\alpha}{n+\beta}-t\biggl{)}^{n-k-1}\\
&=(n+\beta)\{A_{k-1}^{n-1}(t)-A_{k}^{n-1}(t)\}
\end{align*}

which implies

$${\bf{V}}^{\prime}(t)=\sum\limits_{k=0}^{n}~{\bf{P_k}} {(A_{k}^{n}(t)}^{\prime}.$$\\

Now

$${\bf{V^{\prime}}}(\frac{\alpha}{n+\beta})={\bf{P^{\prime}}}(\frac{\alpha}{n+\beta})= (n+\beta)({\bf{P_1}}-{\bf{P_0}})A_{0}^{n-1}(t)$$

and
$${\bf{P^{\prime}}}(\frac{\alpha}{n+\beta})= (n+\beta)({\bf{P_1}}-{\bf{P_0}}){\bigg(\frac{n-1+\beta}{n-1}\bigg)}^{n-1}{\bigg( \frac{n-1+\alpha}{n-1+\beta}-\frac{\alpha}{n+\beta}\bigg)}^{n-1-k}$$

Similarly after some computation, we have
$${\bf{V^{\prime}}}(\frac{n+\alpha}{n+\beta})={\bf{P^{\prime}}}(\frac{n+\alpha}{n+\beta})= (n+\beta)({\bf{P_n}}-{\bf{P_{n-1}}})A_{k-1}^{n-1}(\frac{n+\alpha}{n+\beta}))$$

$${\bf{P^{\prime}}}(\frac{n+\alpha}{n+\beta})= (n+\beta)({\bf{P_n}}-{\bf{P_{n-1}}}){\bigg(\frac{n-1+\beta}{n-1}\bigg)}^{n-1}{\bigg( \frac{n+\alpha}{n+\beta}-\frac{\alpha}{n-1+\beta}\bigg)}^{n-1}$$

 \subsection{Degree elevation for B$\acute{e}$zier curves with shifted knots}

B$\acute{e}$zier curves with shifted knots have a degree elevation algorithm that is similar to that possessed by the classical B$\acute{e}$zier curves.
Using the technique of degree elevation, we can increase the flexibility of a given curve.

 $$ {\bf{ P}}(t) = \sum\limits_{k=0}^{n} {\bf{P_k}}~ G_{n,\alpha,\beta}^k~(t)$$

 $$ {\bf{ P}}(t) = \sum\limits_{k=0}^{n+1} {\bf{P_k^\ast}}~ G_{n+1,\alpha,\beta}^k~(t),$$ where

\begin{equation}\label{e3.7}
{\bf{ P^\ast}_k}=\bigg(1-\frac{n+1-k}{n+1}\bigg)~{\bf{ P_{k-1}}}+ \bigg(\frac{k}{n+1}\bigg)~{\bf{P_k}}
\end{equation}

The statement above can be derived from Theorem \ref{t3.1}. When $ \alpha=\beta=0$ formula \ref{e3.7} reduce to the degree evaluation formula
of the B$\acute{e}$zier curves. If we let $ P = (P_0, P_1, . . . , P_n)^{T}$  denote the vector of control points of the initial  B$\acute{e}$zier
curve of degree $n,$ and $ {\bf{P^{(1)}}}=(P_0^\ast, P_1^\ast, . . . , P_{n+1}^\ast)$
 the vector of control points of the degree elevated B$\acute{e}$zier curve of
degree $n + 1,$ then we can represent the degree elevation procedure as:

$${\bf{P^{(1)}}}=T_{n+1}{\bf{P}},$$
where

$$ T_{n+1}=\frac{1}{n+1}\begin{bmatrix}
\;n+1\; & \;0\; & \;\ldots \; & \;0\; & \;0\; \\
n+1-n & n  & \ldots & 0 & 0 \\
\vdots & \vdots & \ddots & \vdots & \vdots \\
0 & \ldots & n+1 -2  & 2 & 0 \\
0 & 0 & \ldots & n+1-1  & 1 \\
\;0\; & \;0\; & \;\ldots & \;0\; & \;n+1 \;
\end{bmatrix}_{(n+2)\times(n+1)}$$

For any $ l \in \mathbb{N},$ the vector of control points of the degree elevated B$\acute{e}$zier curves of degree $ n + l$ is:
${\bf{P^{(l)}}} = T_{n+l}~ T_{n+2}........ T_{n+1} {\bf{P}}.$
As $l \longrightarrow \infty,$ the control polygon $\bf{P^{(l)}}$ converges to a B$\acute{e}$zier curve.\\


\subsection{de Casteljau algorithm:}

B$\acute{e}$zier curves with shifted knots of degree $n$ can be written as two kinds of linear combination of two B$\acute{e}$zier curves with shifted knots of degree $n-1,$ and we can get the two selectable algorithms to evaluate B$\acute{e}$zier curves with shifted knots. The algorithms can be expressed as:

\textbf{ Algorithm 1.}\\

\begin{equation}\label{e3.8}
\left\{
 \begin{array}{ll}
 {\bf{P^{0}_{i}}}(t)\equiv {\bf{P^{0}_{i}}}\equiv {\bf{P_{i}}}~~~i=0,1,2......,n~~~\mbox{ } &  \\
 &  \\
{\bf{P^{r}_{i}}}(t)=\frac{n+\beta}{n} \bigg(t-\frac{\alpha}{n+\beta}\bigg)~{\bf{P^{r-1}_{i+1}}}(t)+\frac{n+\beta}{n} \bigg( \frac{n+\alpha}{n+\beta}-t \bigg) ~{\bf{P^{r-1}_{i}}}(t)~~~\mbox{  } &\\
 r=1,...,n,~~~i=0,1,2......,n-r.,~~~\frac{\alpha}{n+\beta}\leq t \leq\frac{n+\alpha}{n+\beta},~~0\leq\alpha\leq\beta.~~~~\mbox{  }
 \end{array}%
 \right.
 \end{equation}

Then

 \begin{equation}\label{e3.9}
  {\bf{ P}}(t) = \sum\limits_{i=0}^{n-1} {\bf{P_i^1}}(t)=...=\sum\limits {\bf{P_i^r}}(t)~ b^{i,{n-r}}_{p,q}(t)=...= {\bf{P_0^n}}~(t)
\end{equation}

It is clear that the results can be obtained from Theorem (\ref{t3.2}).  When $\alpha=\beta=0,$ formula (\ref{e3.8}) and (\ref{e3.9}) recover the de Casteljau algorithms of classical B$\acute{e}$zier curves. Let $P^0 = (P_0, P_1, . . . , P_n)^T$ , $P^r = (P_0^r,P_1^r,....,P_{n-r}^r)^{T},$
 then de Casteljau algorithm can be expressed as:\\

\textbf{ Algorithm 2.}

\begin{equation}\label{e3.10}
 {\bf{ P^r}}(t)=M_r(t)....M_2(t)M_1(t){\bf{ P^0}}
\end{equation}
where $M_r(t)$ is a $(n - r + 1) \times (n - r + 2) $ matrix and

$$ M_r(t)=\frac{n+\beta}{n}\begin{bmatrix}
\;\big(\frac{n+\alpha}{n+\beta}-t \big)\; & \;\big(t-\frac{\alpha}{n+\beta}\big)\; & \;\ldots \; & \;0\; & \;0\; \\
0 & \big(\frac{n+\alpha}{n+\beta}-t \big) & \big(t-\frac{\alpha}{n+\beta}\big) & 0 & 0 \\
\vdots & \vdots & \ddots & \vdots & \vdots \\
0 & \ldots & \big(\frac{n+\alpha}{n+\beta}-t \big) &\big(t-\frac{\alpha}{n+\beta}\big) & 0 \\
0 & 0 & \ldots & \big(\frac{n+\alpha}{n+\beta}-t \big) & \big(t-\frac{\alpha}{n+\beta}\big)
\end{bmatrix}$$

\section{Tensor product B$\acute{e}$zier surfaces with shifted knots on $ \big[\frac{\alpha}{n+\beta} , \frac{n+\alpha}{n+\beta}\big]\times  \big[\frac{\alpha}{n+\beta} , \frac{n+\alpha}{n+\beta}\big]$}

We define a two-parameter family ${{\bf{P}}}(u,v)$ of tensor product surfaces of degree $m \times n$ as follow:

\begin{equation}\label{e4.1}
{{\bf{P}}}(u,v) = \sum\limits_{i=0}^{m}\sum\limits_{j=0}^{n} {{\bf{P}}_{i,j}}~G_{m,\alpha,\beta}^i (u)~~ G_{n,\alpha,\beta}^j(v),~~~~(u,v) \in \bigg[\frac{\alpha}{n+\beta} , \frac{n+\alpha}{n+\beta}\bigg]\times  \bigg[\frac{\alpha}{n+\beta} , \frac{n+\alpha}{n+\beta}\bigg],
\end{equation}

where ${{\bf{P}}_{i,j}} \in \mathbb{R}^3 ~~(i = 0, 1, . . . ,m, j = 0, 1, . . . , n),$ and
$G_{m,\alpha,\beta}^i(u),~~  G_{n,\alpha,\beta}^j(v)$ are modified Bernstein functions respectively. We refer to the ${{\bf{P}}_{i,j}}$ as the control points. By joining up adjacent points in the same
row or column to obtain a net which is called the control net of tensor product B$\acute{e}$zier surface.\\

\subsection{Properties}

1. \textbf{Geometric invariance and affine invariance property:} Since
\begin{equation}\label{e4.2}
 \sum\limits_{i=0}^{m}\sum\limits_{j=0}^{n}  G_{m,\alpha,\beta}^i(u)~~ G_{n,\alpha,\beta}^j(v)=1,
\end{equation}

 ${{\bf{P}}}(u,v)$ is an affine combination
of its control points.\\

2. \textbf{Convex hull property:} ${{\bf{P}}}(u,v)$ is a convex combination of ${{\bf{P}}_{i,j}}$ and lies in the convex hull of its control net.\\

3. \textbf{Isoparametric curves property:} The isoparametric curves $v = v^\ast$ and $u = u^\ast$ of a tensor product B$\acute{e}$zier surface are respectively the B$\acute{e}$zier curves with shifted knots of degree $m$ and degree $n,$ namely,

\begin{equation*}
{{\bf{P}}}(u,v^\ast) = \sum\limits_{i=0}^{m}\bigg(\sum\limits_{j=0}^{n} {{\bf{P}}_{i,j}}~G_{n,\alpha,\beta}^j(v^\ast)\bigg)~ G_{m,\alpha,\beta}^i(u)~~ ,~~u  \in \big[\frac{\alpha}{n+\beta} , \frac{n+\alpha}{n+\beta}\big] ;
\end{equation*}

\begin{equation*}
{{\bf{P}}}(u^\ast,v) = \sum\limits_{j=0}^{n}\bigg(\sum\limits_{i=0}^{m} {{\bf{P}}_{i,j}}~G_{n,\alpha,\beta}^j(u^\ast)\bigg)~ G_{m,\alpha,\beta}^i(v)~~ ,~~v  \in \big[\frac{\alpha}{n+\beta} , \frac{n+\alpha}{n+\beta}\big]
\end{equation*}

The boundary curves of ${{\bf{P}}}(u,v)$ are evaluated by ${{\bf{P}}}(u,\frac{\alpha}{n+\alpha})$, ${{\bf{P}}}(u,\frac{n+\alpha}{n+\beta})$, ${{\bf{P}}}(\frac{\alpha}{n+\alpha},v)$ and ${{\bf{P}}}(\frac{n+\alpha}{n+\beta},v)$.\\

4. \textbf{Corner point interpolation property: }The corner control net coincide with the four corners of the surface. Namely, ${{\bf{P}}}(\frac{\alpha}{n+\alpha},\frac{\alpha}{n+\alpha})={{\bf{P}}}_{0,0},$
${{\bf{P}}}(\frac{\alpha}{n+\alpha},\frac{n+\alpha}{n+\beta}) ={{\bf{P}}}_{0,n},$
 ${{\bf{P}}}(\frac{m+\alpha}{m+\beta},\frac{\alpha}{n+\alpha}) ={{\bf{P}}}_{m,0},$
 ${{\bf{P}}}(\frac{m+\alpha}{m+\beta},\frac{n+\alpha}{n+\beta}) ={{\bf{P}}}_{m,n},$\\

5. \textbf{Reducibility:} When $\alpha=\beta=0$ formula (\ref{e4.1}) reduces to a classical tensor product B$\acute{e}$zier patch.

\subsection{Degree elevation and de Casteljau algorithm}
Let ${{\bf{P}}}(u,v)$ be a tensor product B$\acute{e}$zier surface with shifted knots of degree $m \times n.$ As an example, let us take obtaining the same
surface as a surface of degree $(m + 1) \times (n + 1).$ Hence we need to find new control points ${{\bf{P}}}_{i,j}^\ast$ such that

\begin{equation}\label{e4.3}
{{\bf{P}}}(u,v) = \sum\limits_{i=0}^{m}\sum\limits_{j=0}^{n} {{\bf{P}}_{i,j}}~G_{m,\alpha,\beta}^i (u)~~ G_{n,\alpha,\beta}^j(v)= \sum\limits_{i=0}^{m+1}\sum\limits_{j=0}^{n+1} {\bf{P^\ast}_{i,j}}~ G_{m+1,\alpha,\beta}^i(u)~~ G_{n+1,\alpha,\beta}^j(v)
\end{equation}
Let
$\alpha_i=1-\frac{m+1-i}{m+1},~~$ $\beta_j=1-\frac{n+1-j}{n+1}.$

Then
\begin{equation}\label{e4.4}
  {{\bf{P}}}_{i,j}^\ast=\alpha_i~\beta_j~{{\bf{P}}}_{i-1,j-1}+\alpha_i~(1-\beta_j)~{{\bf{P}}}_{i-1,j}+(1-\alpha_i)~(1-\beta_j)~{{\bf{P}}}_{i,j}
\end{equation}

which can be written in matrix form as
$$
  \begin{bmatrix}
    1-\frac{[m+1-i]}{[m+1]} & \frac{[m+1-i]}{[m+1]}\\
  \end{bmatrix}
    \begin{bmatrix}
      {{\bf{P}}}_{i-1,j-1} & {{\bf{P}}}_{i-1,j} \\
     {{\bf{P}}}_{i,j-1} & {{\bf{P}}}_{i,j} \\
    \end{bmatrix}
                   \begin{bmatrix}
                     1-\frac{[n+1-j]}{[n+1]} \\
                     \frac{[n+1-j]}{[n+1]}\\
                   \end{bmatrix}
    $$

The de Casteljau algorithms are also easily extended to evaluate points on a B$\acute{e}$zier surface. Given the control net
${{\bf{P}}}_{i,j} \in \mathbb{R}^3, i = 0, 1, . . . ,m,~~ j = 0, 1, . . . , n.$

\begin{equation}\label{e4.5}
\left\{
 \begin{array}{ll}
 {\bf{P^{0,0}_{i,j}}}(u,v)\equiv {\bf{P^{0,0}_{i,j}}}\equiv {\bf{P_{i,j}}}~~~i=0,1,2......,m;~~j=0,1,2...n.\mbox{ } &  \\
 &  \\
{\bf{P^{r,r}_{i,j}}}(u,v)= \begin{bmatrix}
    \frac{m+\beta}{m}\bigg(\frac{m+\alpha}{m+\beta}-t\bigg)~~~~ & \frac{m+\beta}{m}\bigg(t-\frac{\alpha}{m+\beta}\bigg)\\
  \end{bmatrix}
    \begin{bmatrix}
      {{\bf{P}}}_{i,j}^{r-1,r-1} & {{\bf{P}}}_{i,j+1}^{r-1,r-1} \\
     {{\bf{P}}}_{i+1,j}^{r-1,r-1} & {{\bf{P}}}_{i+1,j+1}^{r-1,r-1} \\
    \end{bmatrix}
                   \begin{bmatrix}
                      \frac{n+\beta}{n}\big(\frac{n+\alpha}{n+\beta}-t\big)~~ \\
                     \frac{n+\beta}{n}\big(t-\frac{\alpha}{n+\beta}\big)~~ \\
                   \end{bmatrix}      ~~~\mbox{  } &\\
 r=1,...,k, ~k=\text{min}(m,n)~~~i=0,1,2......,m-r;~~ j=0,1,....n-r,~~~\frac{\alpha}{n+\beta}\leq t\leq \frac{n+\alpha}{n+\beta}~~\mbox{  }
 \end{array}%
 \right.
 \end{equation}
 or

 When $m = n,$ one can directly use the algorithms above to get a point on the surface. When $m \neq n,$ to get a point on the
surface after $k$ applications of formula (\ref{e4.5}), we perform formula (\ref{e3.10}) for the intermediate point $ {{\bf{P}}}_{i,j}^{k,k}.$\\

\textbf{Note:} We get classical B$\acute{e}$zier curves and surfaces for $(u, v) \in \big[\frac{\alpha}{n+\beta} , \frac{n+\alpha}{n+\beta}\big] \times \big[\frac{\alpha}{n+\beta} , \frac{n+\alpha}{n+\beta}\big] $  when we set the parameter $\alpha=\beta=0.$

\section{ Future work}

 In near future, we will construct $q$-analogue of B$\acute{e}$zier curves and surfaces with shifted knots and we will also study de Casteljau algorithm and degree evaluation properties for curves and surfaces.

\end{document}